\documentclass{aa}  

\usepackage{graphicx}
\usepackage{txfonts}
\begin{document}

%%   \title{Statistical analysis of several hours period intensity pulsations in the solar corona during activity cycle 23}
   \title{Long-period intensity pulsations in the solar corona during activity cycle 23}

   \author{F. Auch\`{e}re \and K. Bocchialini \and J. Solomon \and E. Tison}
	\offprints{F. Auch\`{e}re}

   \institute{Institut d'Astrophysique Spatiale, B\^atiment 121, Univ. Paris Sud - CNRS, 91405 Orsay cedex, France\\
              \email{frederic.auchere@ias.u-psud.fr}
             }

   \date{Received ; accepted }

   \abstract{We report on the detection ($10\ \sigma$) of 917 events of long-period (3 to 16 hours) intensity pulsations in the 19.5 nm passband of the {\it SOHO} Extreme ultraviolet Imaging Telescope. The data set spans from January 1997 to July 2010, i.e the entire solar cycle 23 and the beginning of cycle 24. The events can last for up to six days and have relative amplitudes up to 100\%. About half of the events (54\%) are found to happen in active regions, and 50\% of these have been visually associated with coronal loops. The remaining 46\% are localized in the quiet Sun. We performed a comprehensive analysis of the possible instrumental artifacts and we conclude that the observed signal is of solar origin. We discuss several scenarios which could explain the main characteristics of the active region events. The long periods and the amplitudes observed rule out any explanation in terms of magnetohydrodynamic waves. Thermal nonequilibrium could produce the right periods, but it fails to explain all the observed properties of coronal loops and the spatial coherence of the events. We propose that moderate temporal variations of the heating term in the energy equation, so as to avoid a thermal nonequilibrium state, could be sufficient to explain those long-period intensity pulsations. The large number of detections suggests that these pulsations are common in active regions. This would imply that the measurement of their properties could provide new constraints on the heating mechanisms of coronal loops.}
     
  
   \maketitle
%
%________________________________________________________________

\section{Introduction}

Oscillations in the solar corona have been known for several decades. They occur everywhere, in every type of structure, and their periods can range from a few seconds to several hours. For instance, in solar flares, quasiperiodic pulsations in intensity or in the form of Doppler shifts in the flare spectrum have been detected with a period of a fraction of a second up to several minutes \citep{Dolla12}. They may be due to magnetic reconnections triggered by the kink magnetohydrodynamic (MHD) mode~\citep{Inglis09}. Three-minute period oscillations have been identified in loops related to sunspots and five-minute periods in loops not related to sunspots~\citep{DeMoortel02}, these periods being characteristic of those observed in the chromosphere and the photosphere~\citep{Jensen63}. Various oscillations are observed in active region loops ~\citep{Aschwanden99, Nakariakov99} with a mean period of 5.4 minutes and a damping time  of 9.7 minutes; these oscillations are generated by flares \citep{Hudson04} and interpreted as the kink MHD mode. \citet{Srivastava08} highlighted in cool postflare loops intensity oscillations with periods of 9 min 46 s at their apex and 5 min 49 s near the footpoints. They were attributed to the fundamental and first harmonic of the so-called fast sausage MHD mode. Short-period transverse oscillations (less than 10 minutes), triggered by flares or prominence eruptions, have been detected in coronal loops~\citep{Schrijver02}. In polar plumes, quasi-periodic compressive waves with a 10-15 minute period have been reported~\citep{DeForest98, Ofman99}. \citet{Wang03} detected intensity and velocity oscillations in hot active region loops in \ion{Fe}{XIX} and \ion{Fe}{XXI} lines, with periods ranging from 7 to 31 minutes, and damping times ranging from 5.7 to 36.8 minutes. \citet{Blanco99} and \citet{Bocchialini01} found velocity oscillations in a quiescent prominence simultaneously in four transition region lines, with periods ranging from 6 to 12 minutes. In polar coronal holes, \citet{Banerjee01} found oscillations ranging from 20 to 30 minutes and slow magnetoacoustic waves have been found in intensity and velocity with a dominant period of 25 minutes~\citep{Gupta09}. In bright points, there are intensity oscillations with periods ranging between 8 and 64 minutes~\citep{Tian08}, interpreted in terms of magnetic reconnections. In eruptive filaments, oscillations are clearly observed, in intensity and velocity in the \ion{He}{I} and \ion{Mg}{X} lines, in velocity in H$\alpha$, with similar periods from a few minutes up to 80 minutes, with a main range from 20 to 30 minutes, simultaneously with eruptions \citep{Bocchialini11}. \cite{Zhang12} reported 52-minute period oscillations in an active region prominence observed in the extreme ultraviolet (EUV). In prominences and filaments, three types of periods are measured~\citep{Molowny97}: short period (less than 5 minutes), medium period (6-20 minutes), and long period (40-90 minutes).

In comparison to the abundant bibliography on short and very short oscillations in the solar corona (subsecond to one hour), oscillations of long and very long periods (several hours) have seldom been observed and discussed, perhaps because very long uninterrupted observations are required to detect them. \citet{Pouget06} detected with the Coronal Diagnostic Spectrometer~\citep[CDS;][]{Harrison95} on board the Solar and Heliospheric Observatory~\citep[{\it SOHO};][]{Domingo95}, ultra-long-period oscillations in velocity (5-6 hours) in a filament in the \ion{He}{I}\ 58.4 nm line. Other ultra-long period oscillations in intensity have been detected by~\citet{Foullon04} and \citet{Foullon09}. These oscillations show periods between 8 and 27 hours with a dominant period of 12.1 hours and were attributed to MHD modes in a nearby filament observed in EUV.

These reports have motivated the present work. To find out whether these are isolated cases or a common coronal phenomenon, we systematically looked for these long-period oscillations, not only in filaments but over most of the solar corona, regardless of the localization on the Sun or the type of structure. The aim of this paper is to perform a statistical study of these oscillations to better characterize their properties and understand their physical origin. The database of images taken at 19.5 nm by the {\it SOHO} Extreme ultraviolet Imaging Telescope~\citep[EIT, ][]{delaboudiniere1995} is particularly suited to this analysis because it contains systematic full Sun observations of the corona over a whole solar cycle.

\section{Data processing\label{data}}
\subsection{EIT data over solar cycle 23}
The EIT images are an exceptional database covering more than 17 years of almost uninterrupted observations of the Sun since the launch of {\it SOHO} in December 1995. Among the four wavelengths available, the 19.5 nm passband offers the best cadence with an image every 12 minutes on average. This band is dominated by emission lines of \ion{Fe}{XI} and \ion{Fe}{XII}, with a peak response at $1.6\times 10^{6}$~K. We analyzed data from January 1996 to July 2010 ; after this date the cadence of EIT was changed to 6 hours. The observing program of EIT was generally very stable during this time frame, with the main exceptions of the two intervals when the contact with the satellite was lost, from June to October 1998 and from December 1998 to February 1999. Some time intervals were not processed because the average cadence was longer than 16 minutes. This can be due, for example, to the regular CCD bake-outs (\texttt{http://umbra.nascom.nasa.gov/eit/ bake\_history.html}). The EIT raw data were calibrated using the routine \texttt{eit\_prep} from the Interactive Data Language SolarSoftware library. It corrects for instrumental degradation~\citep{clette2002, benmoussa2013} and returns intensities expressed in $\mathrm{DN}.\mathrm{pix}^{-1}.\mathrm{s}^{-1}$ (DN: Digital Number).

\begin{figure*}
	\centering
		\includegraphics[angle=0, scale=0.9]{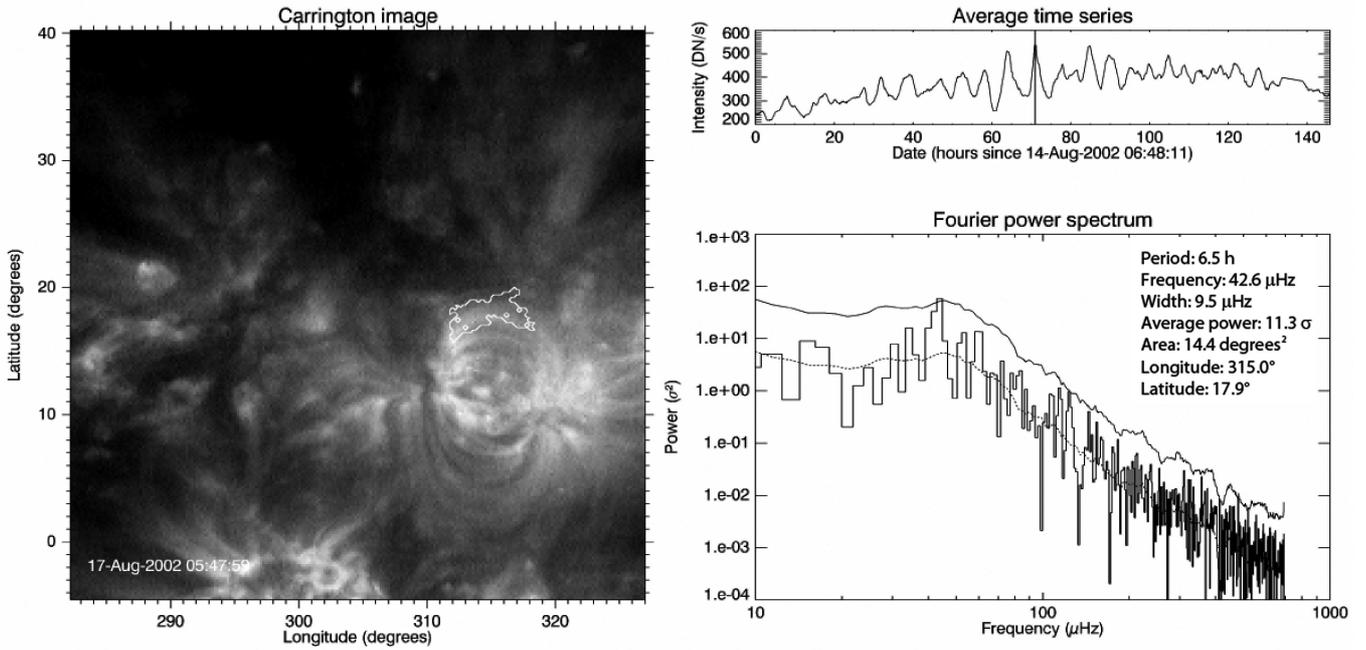}
		\caption{Summary of an event detection in the region of interest tracked starting on August 14, 2002 06:48:11 UT and ending on August 20, 2002 08:36:10 UT. {\it Left}: intermediate frame of the heliographic coordinates movie. The white contour delimits the detected event. {\it Top right}: light curve averaged over the contoured area. The vertical bar marks the date of the frame in the left panel. {\it Bottom right}: Fourier power spectrum of the light curve. The dashed line is the estimate of the background power. The solid line is the 10 $\sigma$ detection threshold. The temporal evolution is available in the on-line edition.}
	\label{fig:oscillations_ar}
\end{figure*}

\begin{figure*}
	\centering
		\includegraphics[angle=0, scale=0.9]{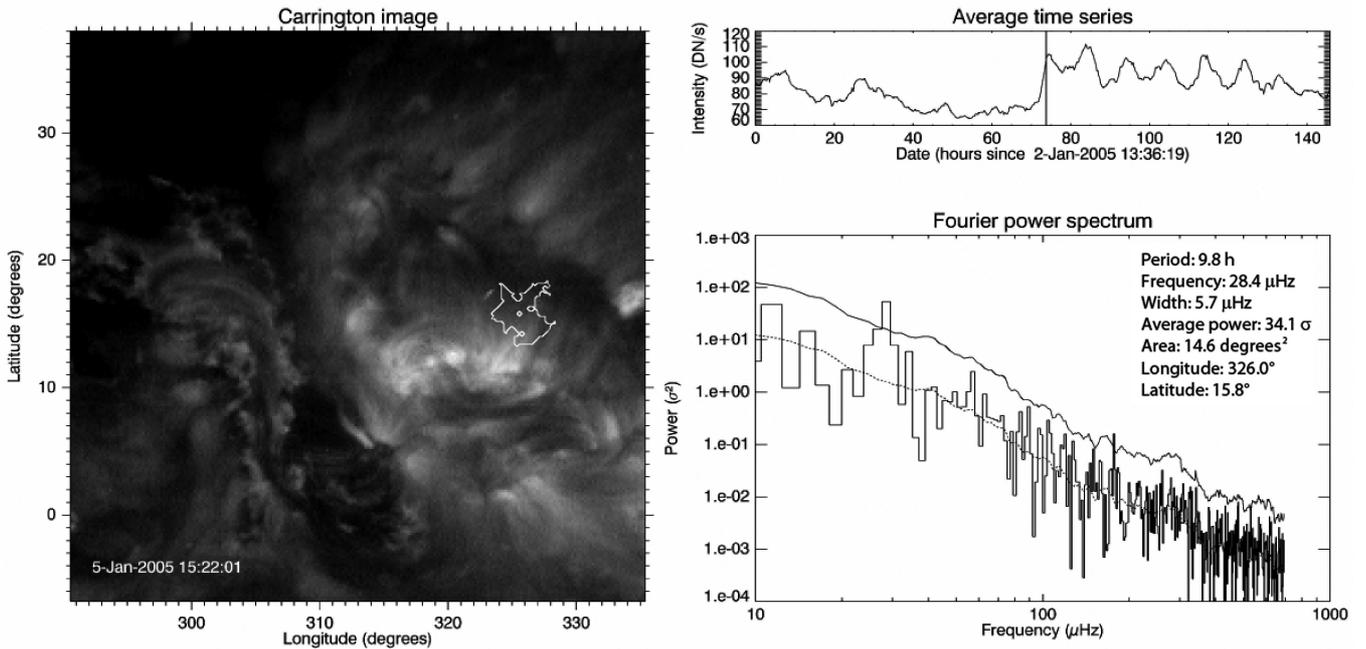}		
		\caption{Same as Fig.~\ref{fig:oscillations_ar} for the sequence running from January 2, 2005 13:36:19 UT to January 8, 2005 15:00:10 UT. The temporal evolution is available in the on-line edition.}
	\label{fig:oscillations_ar_cme}
\end{figure*}

\begin{figure*}
	\centering
		\includegraphics[angle=0, scale=0.9]{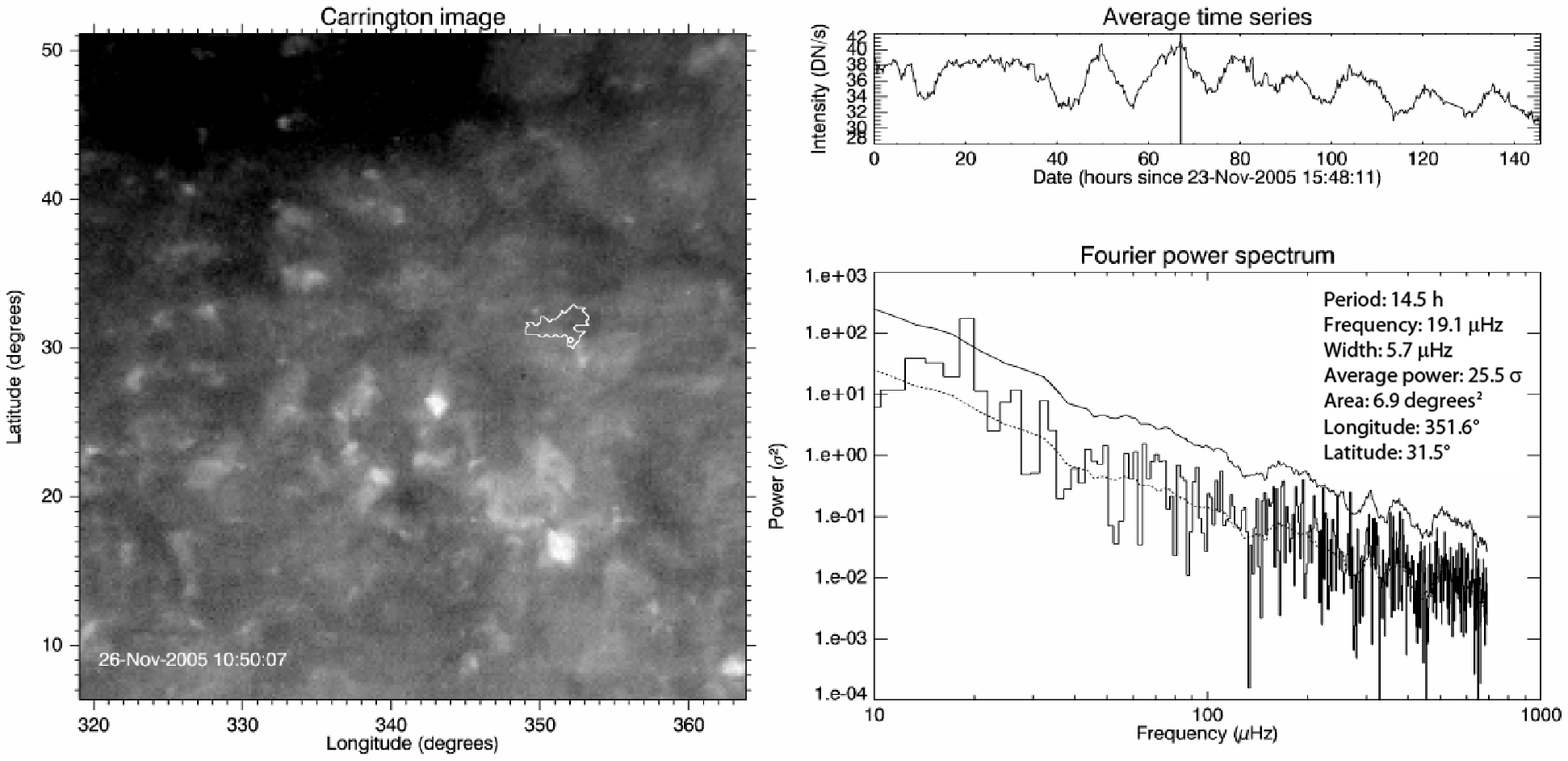}		
		\caption{Same as Fig.~\ref{fig:oscillations_ar} for the sequence running from November 23, 2005 15:48:11 UT to November 29, 2005 17:00:11 UT. The temporal evolution is available in the on-line edition.}
	\label{fig:oscillations_qs}
\end{figure*}

\subsection{Method of detection}
To detect long-period intensity pulsations that last several hours in the solar corona, we used the following four-step automatic method of analysis:
\begin{enumerate}
\item Selection of a region of interest (ROI) on the disk that is tracked as it rotates in order to compute an intensity data cube as a function of the Carrington longitude, latitude, and time.
\item Calculation of the power spectral density (PSD) for every mapped pixel (longitude, latitude) in each ROI.
\item Identification in the PSD cubes (longitude, latitude, frequency) of events matching a pre-defined set of criteria.
\item Storage of the characteristics (area, location, power, frequency, etc.) in a catalogue for statistical analysis.
\end{enumerate}

These four steps are entirely automated, but a final visual examination of the results was performed in order to remove obvious false detections from the catalogue. The details of the analysis are given below. 

For step 1, the ROIs are defined in heliographic coordinates as sectors of $45\degr$ extending $\pm 50\degr$ north and south of the equator. Each ROI is tracked during 146 hours (more than 6 days) from the east limb to the west limb. A new ROI is initiated every 38 hours so that two subsequent ROIs overlap by their half width. The ROIs are tracked by mapping the images into heliographic coordinates. The differential rotation is compensated using the rotation rate $\Omega$ measured by~\cite{Hortin03}
\begin{equation}
\Omega(\phi)=14.51-2.14\times sin^2\phi+0.66\times sin^4\phi\ [\degr.\mathrm{day}^{-1}],
\end{equation}

\noindent
where $\phi$ is the heliographic latitude. This profile is an average for each latitude. We do not take into account that different types of structures can rotate at different rates. Some coronal holes can for example have an almost rigid rotation~\citep{Wang88, Wang93}. We also do not take into account possible meridional flows. The mapping uses a regular 0.2\degr resolution heliographic grid and assumes a solar radius of 696 Mm. Given the date and the position of the SOHO spacecraft, each point of the grid is projected onto the input EIT image plane. Intensities at the resulting coordinates are bi-linearly interpolated between the neighboring data points. Each data cube resulting from these transformations contains about 600 images of $225\times 500$ heliographic pixels.

For step 2, we compute the PSD for each time series corresponding to each heliographic pixel. Since the EIT cadence is not exactly stable, we re-sample each time series on a regular grid before transformation to the Fourier space. The re-sampling frequency is twice the median of the data acquisition frequency, i.e., around six minutes for most of the cubes. The times series are apodized with a sine at both ends on 5\% of the signal duration to avoid aliasing. The variance normalized PSD for every pixel is then computed using a fast Fourier transform. The typical frequency resolution of the PSD cube is $2\ \mu \mathrm{Hz}$. It is worth noting that no detrending of the time series is applied before computing the PSD. Such detrending, for example by removing a running boxcar average, can bias the interpretation of the PSD. Indeed, since the PSDs tend to follow a power law, and since detrending suppresses the low frequencies, the PSDs of detrended series tend to show a hump at frequencies just above the cut off of the boxcar that can be incorrectly interpreted as a signature of dominant frequencies in the original signal.

In step 3, we scan each resulting PSD cube for events using the following procedure. The PSD at each heliographic location is treated independently from the others. In order to identify frequencies where a coherent signal is present, each frequency bin of the PSD is compared to the local mean power, assumed to be locally white Gaussian noise. In that case, the probability that a frequency bin of the PSD has a value greater than $m$ times the local mean~$\sigma$ is given by $1 - (1 - e^{-m})^N$~\citep{Gabriel02}, with $N$ the number of points of the PSD (typically 350). We set the detection level to be 10 times the local background estimated using a smoothed copy of the PSD, which corresponds to a confidence level of 99\% for individual PSDs. However, since we analyze $225\times 500$ locations, we actually have $N\approx 4\times10^7$, and the probability of having at least one random peak of power above $10\ \sigma$ in the full ROI is practically equal to one. But the probability that several pixels are adjacent and simultaneously above threshold drops very rapidly with the number of pixels. Therefore, we cluster 6-connected PSD voxels (neighbors touching on one of their faces along the spatial or the frequency direction) above the $10\ \sigma$ threshold, and we discard the resulting regions having an area smaller than 2 square degrees (50 heliographic pixels), which is the typical area of bright points observed by EIT. We also discard regions that are only one heliographic pixel wide in latitude or longitude, and we impose that the peak of power does not spread over more than $10\ \mu \mathrm{Hz}$. Finally, in order to have at least ten data points per period or ten periods in the signal duration, we consider only frequencies between $18\ \mu \mathrm{Hz}$ and $185\ \mu \mathrm{Hz}$. Each one of the remaining clusters is then grown from its peak of power until the $5\ \sigma$ level is reached.

In step 4, the main characteristics of the detected regions are computed and stored in an ASCII file for further analysis: date, average heliographic longitude and latitude, surface area, maximum power, average power, average frequency, frequency width, average light curve and corresponding PSD. A summary movie is also produced for each event. 

\section{Results and statistical analysis\label{sec:statistics}}

\subsection{Sample events} 

Fig.~\ref{fig:oscillations_ar} gives an example of an intermediate frame of one of the summary movies\footnote{The three movies corresponding to the examples in Figs.~\ref{fig:oscillations_ar},~\ref{fig:oscillations_ar_cme}, and~\ref{fig:oscillations_qs} are available at \url{ftp://ftp.ias.u-psud.fr/fauchere/pub/pulsations/}. Other cases are available upon request from the authors.}. The EIT sequence started on August 14, 2002 06:48:11 UT and ended on August 20, 2002 08:36:10 UT. The left panel displays the ROI in heliographic coordinates at the middle of the sequence ; the observation date is noted at the bottom left. The detected event is outlined by the white contour centered on 315.0\degr\ of longitude and 17.9\degr\ of latitude. It has an area of 14.4 heliographic square degrees (i.e., $1669\ \mathrm{Mm}^2$). As in 29\% of cases (see Sect.~\ref{sec:statistics}), this contour seems to trace underlying coronal loops, a property that will be used later in the analysis. The top right panel shows the light curve averaged over the contoured region. The vertical bar marks the date of the heliographic image of the left panel. Pulsating intensity variations are visible starting from 25 hours after the beginning of the sequence up to 15 hours before its end. The variations temporarily reach a relative amplitude of 100\% (250 to 500\ $\mathrm{DN.s}^{-1}$) around the middle of the sequence. The bottom right panel shows the corresponding Fourier power spectrum, normalized to the variance $\sigma_0^2$ of the light curve~\citep{Torrence1998}. The dashed line represents the estimate of the background power, and the solid line is the $10\ \sigma$ detection threshold. The power spectrum follows a power law from the highest frequencies down to $50\ \mu$Hz and then flattens out at low frequencies. The peak of power is centered on $42.6\ \mu$Hz, i.e., 6.5\ hours, and is spread over three to four frequency resolution bins. The power at the peak is $11.3\ \sigma$. 

Fig.~\ref{fig:oscillations_ar_cme} shows another example of a detection in a small active region. The sequence runs from January 2, 2005 13:36:19 UT to January 8, 2005 15:00:10 UT. This time the contour delimits an area of 14.6 heliographic square degrees ($1692\ \mathrm{Mm}^2$) centered on 326.0\degr\ of longitude and 15.8\degr\ of latitude. The contour is not shaped as a loop, but the underlying features are clearly the base of coronal loops. The left frame corresponds to the starting time of the pulsations visible in the second half of the average light curve. The power spectrum follows a power law over the whole range of frequencies and there is a $35.4\ \sigma$ peak of power at $28.4\ \mu$Hz (9.8 hours). Interestingly, a coronal mass ejection (CME) occurred between 13:45 UT and 15:22 UT (8 EIT frames are missing) from roughly 305\degr\ of longitude and 15\degr\ of latitude. It is listed as a fast (831\ $\mathrm{km.s}^{-1}$ at 20\ $R_\sun$) full halo CME (360\degr) in the catalogue of~\citet{gopalswamy09}. At 15:22 UT, the intensity sharply increased by 40\%, followed by 15-20\% amplitude pulsations until the end of the sequence. The fact that the start of the oscillations closely follows the CME may be a coincidence, but causality is also a possibility. Indeed, as discussed in section~\ref{sec:interpretation}, the observed pulsations may be due to cyclical variations of the temperature and density of loops around their equilibrium state. The propagation of the CME front may have modified the physical conditions in this regions sufficiently to trigger a series of cyclical variations of the plasma parameters from an initial state of equilibrium. Other examples and further investigation is, of course, needed to confirm this hypothesis.

Fig.~\ref{fig:oscillations_qs} shows a last example of a detection in a quiet Sun region. The sequence started on November 23, 2005 15:48:11 UT and ended on November 29, 2005 17:00:11 UT. This is a smaller region of $800\ \mathrm{Mm}^2$ centered on 351.6\degr\ of longitude and 31.5\degr\ of latitude. The power spectrum follows a power law and has a $25.5\ \sigma$ peak of power at $19.1\ \mu$Hz (14.5 hours). The pulsations last for about two thirds of the sequence and have an amplitude of up to 20\%.

\subsection{Statistical properties\label{sec:statistics}} 

\begin{figure*}
\centering
\includegraphics[scale=1.0]{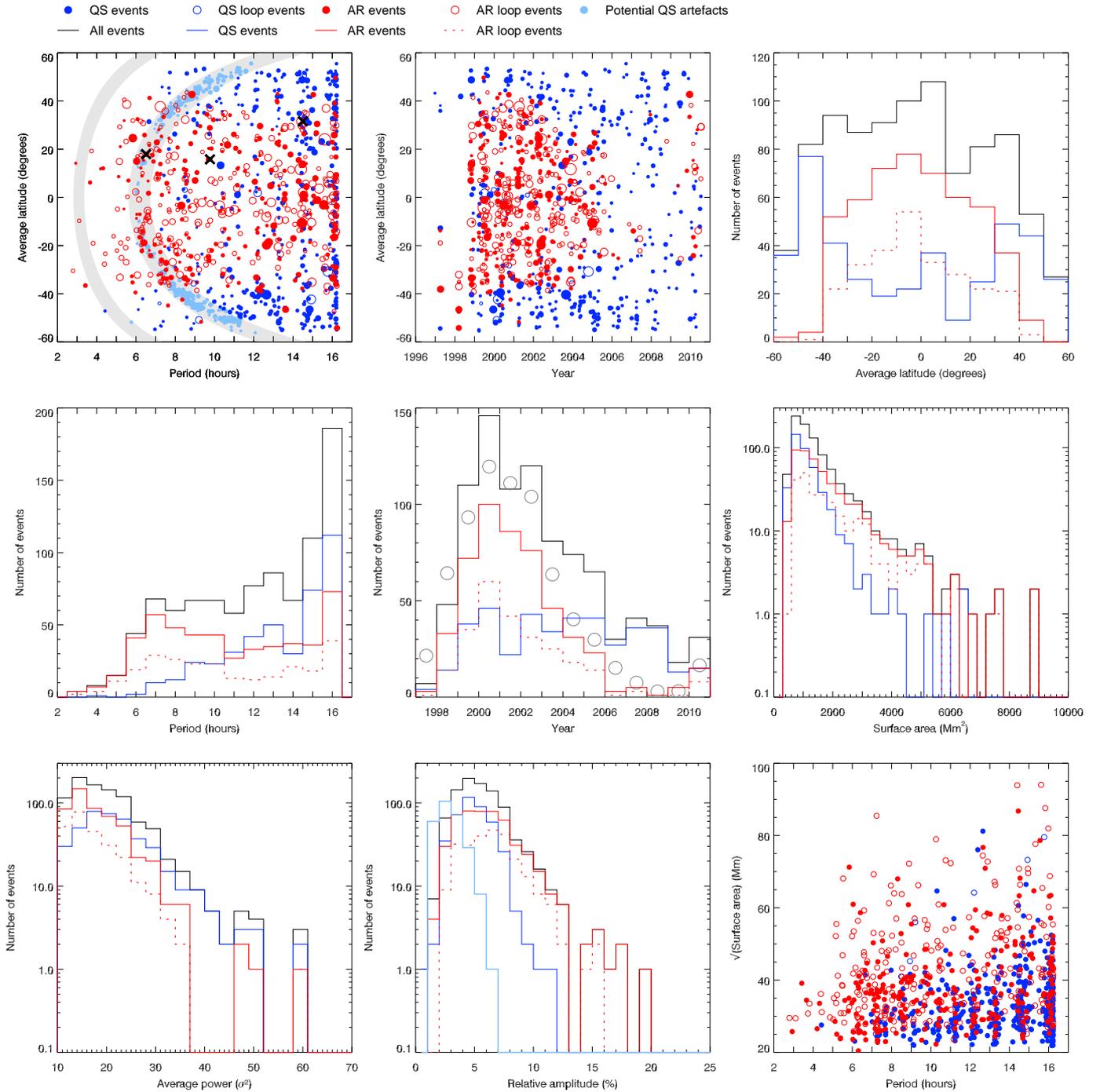}
\caption{Summary of the main properties of the detected events. {\bf Top left}: latitude of the events {\it vs.} their period. In this representation, artifacts can occur in the two shaded regions (see Appendix~\ref{sec:biases} and Fig.~\ref{fig:simul}). Quiet Sun (QS) events clustered in these regions (in light blue) are potential artifacts and are excluded in the subsequent plots. Other QS events are denoted by blue circles, and Active Region (AR) events by red circles. Open circles correspond to events associated with loops. The area of the circles is proportional to the heliographic area of the events. From left to right, the black crosses mark the events illustrated in Figs.~\ref{fig:oscillations_ar},~\ref{fig:oscillations_ar_cme}, and~\ref{fig:oscillations_qs}, respectively. {\bf Top center}: latitude {\it vs.} date. QS and AR events outline mutually exclusive areas and the AR events draw a butterfly-type pattern. {\bf Top right}: Number of events as a function of latitude. The global distribution (black) is rather flat and results from the opposite behavior of QS and AR events. {\bf Center left}: histograms of periods. if one considers all events (black curve), the longer the period the more detections, with a plateau between 7 and 14 hours. AR events (red) dominate the short periods, and after a hump around 7 hours the distribution is flat. AR events associated with loops have the same behavior. QS events (blue) become more frequent with increasing periods. {\bf Center}: number of events per year. The number of QS events is about constant while the number of AR events closely follows the Yearly Sunspot Number (gray circles). {\bf Center right}: histograms of areas, in log scale. QS and AR events follow exponential distributions, with AR events tending to be larger. {\bf Bottom left}: Histograms of the power of events. All types of events roughly follow exponential distributions of similar slopes. {\bf Bottom center}: Histograms of amplitudes relative to average light curve intensities. AR events (solid and dashed red) tend to have larger amplitudes than QS events (blue). The smaller amplitudes of the potential artefacts identified in the top left panel (light blue histogram) clearly set them apart from  other events. {\it \bf Bottom right}: square root of the events area {\it vs.} their period. No correlation is apparent, but there is a region of exclusion: events with short periods tend to have smaller areas.}
\label{fig:synthesis_histograms}
\end{figure*}

The code returned a total of 1236 detections. As a sanity check, we visually examined a summary of every detection in movie format. We thus checked for duplicate events (the same events detected in subsequent data series) and we identified obvious false detections (e.g., due to corrupted data). Each event was also classified as being located either in an active region (AR), in the quiet Sun (QS), or in another type of region. This classification was done visually and can therefore be affected by subjective biases. The events in Figs.~\ref{fig:oscillations_ar} and~\ref{fig:oscillations_ar_cme} were labeled AR, while the event in Fig.~\ref{fig:oscillations_qs} was labeled QS. Since some regions seemed to have the shape of a loop (see Fig.~\ref{fig:oscillations_ar}), the events were also tagged if their contour outlined a portion of visible loops at any time during the sequence. The event in Fig.~\ref{fig:oscillations_ar} was thus tagged as associated with coronal loops.

As demonstrated in Appendix~\ref{sec:biases}, spurious periods can be present in the EIT data in precise regions of the latitude {\it vs.} period space (see Fig.~\ref{fig:simul}). The distribution of the detected events in this space is shown in the top left panel of Fig.~\ref{fig:synthesis_histograms}. The two shaded areas correspond to the possible instrumental periods computed using Equ.~\ref{eq:artifacts} assuming average Stonyhurst longitudes $\theta$ in the range [17\degr, 37\degr]. The blue circles indicate events associated with QS and red circles indicate events associated with ARs. Open circles denote events associated with loops. The area of the circles is proportional to the heliographic surface area of the events. It is apparent that the AR events are evenly distributed, while many of the QS events cluster in the right shaded area. These 319 events, marked in light blue, are therefore potential artifacts. Their relative amplitudes are given by the light blue histogram in the bottom center panel in Fig.~\ref{fig:synthesis_histograms}. Compared to the other events, they have smaller amplitudes compatible with what is expected for artifacts (see Appendix~\ref{sec:biases}), which confirms that they form a distinct population. Since the AR events do not exhibit a clear clustering in the latitude {\it vs.} period space, all of them were kept even if located in the shaded areas. The three events in Figs.~\ref{fig:oscillations_ar},~\ref{fig:oscillations_ar_cme}, and~\ref{fig:oscillations_qs} are marked by black crosses. The event in Fig.~\ref{fig:oscillations_ar} is the leftmost cross. It is in the middle of the right shaded area and is thus a potential artifact. However, the amplitude of the pulsations in this event (locally up to 100\%) is at least an order of magnitude larger than what is expected for instrumental artifacts (a few percent, see Appendix~\ref{sec:artifacts}). This is a common property that justifies the choice to keep the AR events located in the artifact regions. The two other events in Figs.~\ref{fig:oscillations_ar_cme} and~\ref{fig:oscillations_qs} are completely outside the shaded areas and are therefore not artifacts. 

After the identification of duplicates, corrupted sequences, and potential artifacts, 917 valid events were included in the statistical analysis. From the visual labeling of the environment of the events, 411 (45\%) are located in the quiet Sun, 499 (54\%) in active regions, and 7 (1\%) in other types of structures. Those associated with ARs are in general located in the outer part of the region, not in the core. Of the 268 events (29\%) that were tagged as being associated with coronal loops, 95\% (254) are located in active regions, which represents 51\% of AR events. The remaining 13 (5\%) events located in the quiet Sun may be associated with large loops interconnecting two active regions. 

The center left panel in Fig.~\ref{fig:synthesis_histograms} gives the histogram of the periods of all the events (black), of the QS events (blue), of the AR events (red), and of the AR events tagged as associated with loops (dashed red). It can be thought of as the projection of the top left panel on the periods axis. Globally, the number of detection increases with the period, with a flat between 7 and 14 hours. The trend is more pronounced for QS events. Only one of them has a period shorter than 6 hours and there is a steady increase of the number of detections with increasing period. The median period is 14 hours. The distribution is different for AR events, with a median at 10 hours. AR events dominate the short periods, with a hump around 7 hours and a rather flat distribution above 10 hours. AR loop events follow the same kind of distribution. Since we did not exclude any AR events, the hump may be due to a few artificial detections located in the 6-8 hours range of the area of potential artifacts (see top left panel). This has to be investigated further using a more sophisticated discrimination of artifacts. Further differences between QS and AR events are visible when looking at the temporal evolution. The top center panel in Fig.~\ref{fig:synthesis_histograms} is a scatter plot of the average latitude of the events {\it vs.} the average date of the corresponding data sequence. The color and size coding are the same as in the top left panel. The QS and AR events are spread in mutually exclusive regions and the AR events draw a butterfly pattern. The central panel shows the histogram of events per year. It can be thought of as the projection of the top center panel on the dates axis. The distribution for all events (black) has two peaks in 2000 and 2002. The number of QS events (blue) is almost constant with time. The reduced number of events from 1996 to 1998 can be explained by extended periods with insufficient data cadence in 1996-97, and the total interruption of the {\it SOHO} mission for several months in 1998. The number of AR (solid red) and AR loop events (dashed red) closely correlate with the Yearly Sunspot Number (gray circles) maintained at the Solar Influences Data Analysis Center~\citep{sidc}. The top right panel shows no preferential latitude when all the events are considered. The detections are slightly more frequent in the Southern Hemisphere (55\%), which is consistent with the asymmetry of activity during most of cycle 23~\citep{Temmer2006, Auchere2005}. The AR events are dominant at equatorial and mid-latitudes, while the QS events are dominant at high latitudes. The center right panel shows that the distribution of the areas of both QS and AR events follow exponential distributions, but with different indices. The slope of the distribution of QS events is steeper than that of AR events, which reflects a tendency of the QS events to be smaller than the AR events. The distributions of the power of AR and QS events are very similar (bottom left panel). The bottom center panel gives the histograms of the amplitudes of the dominant Fourier components relative to the average intensities of the light curves. We note that an event can temporarily have an amplitude much larger than that of its main Fourier component (see, e.g., Fig.~\ref{fig:oscillations_ar}). The AR events (red) tend to have larger relative amplitudes than the QS events (blue). The bottom right panel is discussed at the end of section~\ref{sec:interpretation}.

Although it would be an interesting characteristic to compare other parameters with, we do not have measurements of the duration of the events. This would require a different type of analysis, for example using wavelet instead of Fourier transforms  in order to preserve the temporal information. Anyway, since the number of days for which we can track a ROI is limited, we do not see the beginning or the end of every event, which will inevitably affect the statistical analysis of their duration. However, from visual inspection, it appears that many events last several tens of hours, and up to 100 hours or more for some of them. 

\section{Physical interpretation\label{sec:interpretation}}
The statistical analysis of the results has shown that events located in the quiet Sun have different properties than those located in active regions. This suggests that the two classes may have different physical origins. We have not yet identified a pattern that could guide us towards an understanding of QS events. In the case of AR events however, 51\% of them have been visually identified as associated with coronal loops. In addition, it is likely that more events are associated with loops. In many cases the loops or loop bundles in active regions are not clearly identifiable. This can be the case if only the base of the loops is visible (see Fig.~\ref{fig:oscillations_ar_cme}), or if several structures are superimposed along the line of sight. Therefore, we do not expect all the areas of significant power to be shaped as loops, which was our visual criterion. Thus, we will focus in this section on the interpretation of the AR events, assuming that a majority of them are intensity pulsations in coronal loops. 

Intensity variations of long periods (10 to 30 hours) have been detected previously by~\citet{Foullon04, Foullon09} in EIT 19.5 nm sequences. These authors have attributed these phenomena to filaments viewed in absorption in the EUV. Other occurrences of long periods in filaments have been reported. In the specific case of quiescent filaments, Fourier analysis of the Doppler velocities in very long time sequences of the {\it SOHO} spectrometers have yielded detections of oscillations with periods of up to 5 hours~\citep{Regnier01, Pouget06}. These observations were interpreted using the prominence model of~\citet{Joarder1993} to provide diagnostics of these structures. It is clear that in filaments, the relatively low temperature of the plasma ($T \leq {10^{4}}$~K) results in low values of the sound speed ($C_s = 1.5\times10^2 \sqrt{T}\approx{15}$~km.s$^{-1}$) allowing, in principle, slow waves with periods of several hours to exist~\citep{Joarder1992, Roberts00}. However, as we have seen, the visual examination of hundreds of cases have led us to conclude that most of the active region events detected in the 19.5 nm band of EIT are linked to coronal loops, not to filaments. For a coronal structure, a slow mode period $\tau\approx{7}$~hours would imply a length $L\approx{2400}$~Mm for a temperature $T\approx 1.6\times 10^6$~K ($C_s\approx 190\ \mathrm{km.s}^{-1}$). This is three times longer than some of the longest reported loop lengths ~\citep[$\approx 800$ Mm, e.g., ][]{Robbrecht01}, which precludes any explanation in terms of slow waves. It is possible to imagine that coupling mechanisms in systems of coronal loops in complex active regions could give rise to large periods by some beat effects, but this does not seem to emerge from recent studies~\citep{VanDoorsselaere08, Luna09}. 

Thermal nonequilibrium is a mechanism that can produce periods compatible with the observed ones. In that case, the periodical intensity variations are not a signature of modes of oscillation but they are the result of a condensation cycle of the coronal plasma in loops~\citep[e.g.,][]{Karpen2001}. The authors shown that asymmetric heating of the footpoints of a loop leads to a cycle of condensation, drift, and destruction of the plasma. In their simulations this cycle has a period of 7 to 10 hours, i.e., well in the range of periods measured in this work (Fig.~\ref{fig:synthesis_histograms}). Numerical simulations of thermal nonequilibrium were also performed by~\citet{Muller05} to explain the observation of coronal rain, i.e., the creation at the top of a coronal loop of a region of high density plasma (blob of plasma) which then falls down because of gravity (see, e.g., their Fig. 2). But while these studies consider thermal nonequilibrium as a way to form cool plasma blobs and prominences by condensation,~\citet{Mok08} have proposed that thermal nonequilibrium could also explain the formation of coronal loops. These authors solved the plasma dynamics parallel to static field lines, using a three-dimensional magnetic field extrapolated from {\it SOHO}/MDI magnetograms. For high enough values of the local heating rate $H$, supposed to be concentrated at the footpoints of the loops, the solution becomes thermally unstable: individual loops evolve through large amplitude condensation-evaporation cycles and the temperature oscillates in time with a period of about 16 hours (see their Figure 2). 
 
Nevertheless, inconsistencies exist between observations and the results of numerical simulations.~\citet{Klimchuk10} examined critically whether thermal nonequilibrium can reproduce the properties of observed coronal loops by performing one-dimensional simulations with a steady coronal heating decreasing exponentially with height and concentrated near both footpoints (and asymmetric with respect to these footpoints). In the case of monolithic loops, emissions in the simulations appeared too localized with respect to the observations. In the case of multi-stranded loops the simulated emission is more uniform, but the loops are then too long-lived. In any case, considering the size of the regions that we detected, many strands and eventually many loops would need to evolve in phase to reproduce the observed spatio-temporal coherence. This imposes a strong constraint on the mechanism responsible for these phenomena.

Given these problems, we went back to considering the simple situation of a one-dimensional constant pressure loop ($P=P_0$) in order to investigate if periods could occur naturally without having to resort to thermal nonequilibrium. One assumes that $z_\mathrm{apex} << \Lambda\approx 50\ T\sim 80$ Mm, where $z_\mathrm{apex}$ is the altitude above the photosphere of the apex of the loop and $\Lambda$ is the pressure scale-height with $T\sim T_\mathrm{max}\sim 1.6\times 10^6$ K. Under these hypotheses, and neglecting flows, the energy conservation equation can be written as~\citep[e.g.,][]{Priest82}
\begin{equation}
\rho C_p \frac{\partial T}{\partial t}=E_h-P_0^2\chi_0T^{-(2+\gamma)}+\frac{\partial}{\partial s}\left(\kappa_0T^{5/2}\frac{\partial T}{\partial s}\right),
\label{eq:energy}
\end{equation}
with $s$ the distance along a loop of total length $2L$, $t$ the time, $T(s, t)$ the temperature, $\rho(s, t)$ the mass density along the loop, and $C_p\approx 2.07\times 10^4\ \mathrm{J.K^{-1}.kg^{-1}}$ the specific heat at constant pressure. The three terms of the right member of this equation are the heating term, then the radiative losses term, and the heat conduction term. 
One has $P=P_0=2n_\mathrm{e}k_\mathrm{B}T\approx 2\rho k_\mathrm{B}T/m_\mathrm{p}$, with $n_\mathrm{e}$ the electron density, $k_\mathrm{B}$ the Boltzmann constant, and $m_\mathrm{p}$ the proton mass. At coronal temperatures (we take $T\sim1.6\times 10^6$~K) a good approximation for the radiative losses term is obtained with $\gamma=0.5$ and $\chi_0\approx 2.6\times 10^{13}\ \mathrm{m^3.J^{-1}.s^{-1}}$~\citep{Martens10}. For the heat conduction term one has $\kappa_0\approx 10^{-11}\ \mathrm{J.m^{-1}.s^{-1}}$. 
We will not discuss the different parametrizations adopted by different authors for the heating term $E_h$, with various dependences on $s$ and eventually on $t$. For the purpose of our discussion, we only need to use a rough space averaged value $\bar{E_h}$ of $E_h$: $\bar{E_h}\sim E_h\times \lambda$ where $\lambda$ is the characteristic scale height of the energy deposition along the loop.

We first consider briefly the static case by equating the right member of Equ.~\ref{eq:energy} to zero. By solving this last equation one can get an expression for $T(s)$ which depends on $T_\mathrm{max}$, the maximum temperature at the apex of the loop ~\citep[for a recent discussion of this static problem, see, e.g.,][]{Martens10}. In the spirit of the classical RTV scaling laws~\citep{Rosner78}, one also gets two relations between $P_0$, $L$, $T_\mathrm{max}$, and $E_h$
\begin{equation}
P_0L\approx 3\times 10^{-13}\ T_\mathrm{max}^3\ \mathrm{J.m^{-2}}
\label{eq:static}
\end{equation}
\noindent
and (for uniform heating)
\begin{equation}
E_h\approx3\times 10^3\ P_0^{7/6}L^{-5/6}\ \mathrm{W.m^{-3}}.
\end{equation}

\noindent
Thus, for given values of $T_\mathrm{max}$ and $L$, the value of the heating term $E_h$ is fixed. Yet, it is worthwhile to note that the upper part of the temperature profile $T(s)$ becomes increasingly flatter when changing from top heating to uniform heating to footpoint heating. Eliminating $P_0$ between the last two expressions, one gets
\begin{equation}
E_h\approx 9\times 10^{-12}\ T_\mathrm{max}^{7/2}L^{-2}\ \mathrm{W.m^{-3}}.
\end{equation}

\noindent
Taking $T_\mathrm{max}=1.6\times 10^6$~K and $L=50$~Mm, one has $E_h\approx 1.8\times 10^{-5}\ \mathrm{W.m^{-3}}$. For the case of uniform heating ($\lambda=L$) one obtains for the static equilibrium state $\bar{E_h^\mathrm{e}}\approx 1.8\times 10^{-5}\times L\approx 9\times 10^2\ \mathrm{W.m^{-2}}$. In the following, exponents $^\mathrm{e}$ and $^\mathrm{ne}$ are used to differentiate the static equilibrium energy $\bar{E_h^\mathrm{e}}$ from the energy of the system $\bar{E_h^\mathrm{ne}}$ in another state (e.g., thermal nonequilibrium). Characteristic values of $\bar{E_h}$ used by~\citet{Muller05} and by~\citet{Klimchuk10} in their numerical simulations of thermal nonequilibrium (for comparable values of $T_\mathrm{max}$) are $\bar{E_h^\mathrm{ne}}\approx 5\times 10^3-10^4\ \mathrm{W.m^{-2}}$ (roughly the average coronal energy losses from an active region), that is about one order of magnitude larger than the value corresponding to the static equilibrium state.
In the time dependent case, small enough temporal perturbations of the heating term $\bar{E_h}$ such that $\bar{E_h^\mathrm{e}}<\bar{E_h}<\bar{E_h^\mathrm{ne}}$ could yield periodic variations while avoiding thermal nonequilibrium conditions. Using Equ.~\ref{eq:energy} we estimate the characteristic time scales $\tau_\mathrm{rad}$ of the radiative losses term and $\tau_\mathrm{cond}$ of the heat conduction term. From this equation, one gets (with $n_\mathrm{e}$ in m$^{-3}$)
\begin{equation}
\tau_\mathrm{rad}(s)\sim \frac{\rho C_pT^{3+\gamma}}{P_0^2\chi_0}\sim \frac{m_pC_pT^{1+\gamma}}{4n_ek_B^2\chi_0}\sim 2\times 10^9\frac{T^{3/2}}{n_\mathrm{e}},
\label{eq:tau_rad}
\end{equation}
and
\begin{equation}
\tau_\mathrm{cond}(s)\sim\frac{\rho C_pL^2}{\kappa_0T^{5/2}}\sim 4\times 10^{-12}\ \frac{n_\mathrm{e}L^2}{T^{5/2}}.
\label{eq:tau_cond}
\end{equation}

We consider the case where $\tau_\mathrm{rad}\approx \tau_\mathrm{cond}$, that is when heat conduction and radiative losses are comparable. From Equs.~\ref{eq:tau_rad} and~\ref{eq:tau_cond}, one obtains an estimate $n_\mathrm{e}^*$ of the corresponding density
\begin{equation}
n_\mathrm{e}^*=2.25\times 10^{10}\frac{T_\mathrm{max}^2}{L}
\label{eq:nestar}
\end{equation}

In Equ.~\ref{eq:nestar}, we choose to take $T\sim T_\mathrm{max}$. We note that this last approximation should not be too critical at least for those loops that are measured to be roughly isothermal~\citep[e.g.,][]{Aschwanden01}. One can compare the typical density $n_e^*$ with the minimum value of the density at the apex obtained by using the static equilibrium Equ.~\ref{eq:static} with $T=T_\mathrm{max}$:
\begin{equation}
n_\mathrm{e}^{\mathrm{apex}}\approx 1.1\times 10^{10}\frac{T_\mathrm{max}^2}{L}.
\end{equation}

\noindent
This results in $n_\mathrm{e}^*/n_\mathrm{e}^\mathrm{apex}\approx 2.1$, i.e., a gravitationally stable configuration. With $T_\mathrm{max}=1.6\times 10^6$~K and $L=50$~Mm, one has $n_\mathrm{e}^*\approx 1.15\times 10^{15}\ \mathrm{m^{-3}}$ and $\tau_\mathrm{rad}\approx\tau_\mathrm{cond}\approx 1\ \mathrm{h}$. This last result gives a first indication of the possible time scale of the evolution of the system. At this point, some words of caution are needed: 
\begin{itemize}
\item Relaxing the constraint $P=P_0=\mathrm{constant}$ along the loop would change $n_e(s)$ and would modify the expression obtained for $n_\mathrm{e}^*$ (Equ.~\ref{eq:nestar}), particularly for the longer loops~\citep{Serio81, Aschwanden01}.                                                                                                                                                                                                                                                                                                                                                                                                                                                                                                                                                                                                                                    \item Observations show that one often departs from the situation where $\tau_\mathrm{rad}\approx\tau_\mathrm{cond}$, with either $\tau_\mathrm{rad}<\tau_\mathrm{cond}$ (overdense case) or $\tau_\mathrm{rad}>\tau_\mathrm{cond}$ (underdense case)~\citep[e.g.,][]{Klimchuk06}.
\end{itemize}

Keeping these caveats in mind, using Equ.~\ref{eq:nestar} to substitute $n_\mathrm{e}^*$ in Equ.~\ref{eq:tau_rad} or~\ref{eq:tau_cond}, one gets (for the common value of $\tau_\mathrm{rad}$ and $\tau_\mathrm{cond}$) a characteristic time
\begin{equation}
\tau_\mathrm{rc}\sim 9\times 10^{-2}\frac{L}{\sqrt{T_\mathrm{max}}}
\label{eq:tau_rc}
\end{equation}

%\begin{figure}
%\centering
%\includegraphics[scale=0.6]{data-Figures/size_vs_period.eps}
%\caption{Square root of the events area {\it vs.} their period. Blue circles are QS events, red circles are AR events. Open circles correspond to events associated with loops. No correlation is apparent, but there is a region of exclusion: events with short periods tend to have smaller areas.}
%\label{fig:size_vs_period}
%\end{figure}

\noindent 
that is $0.2\ \mathrm{h}\leqslant \tau_\mathrm{rc}^\mathrm{apex}\leqslant 6\ \mathrm{h}$ for $T_\mathrm{max}=1.6\times 10^6$~K and $10\ \mathrm{Mm}\leqslant L\leqslant 300\ \mathrm{Mm}$. 
There remains the question of the characteristic excitation times $\tau_\mathrm{exc}$ of the unknown heating processes. To obtain pseudo-periodic fluctuations such as in our observations, one should have $\tau_\mathrm{exc}<\tau_\mathrm{rc}$. Therefore, one can infer that a small enough temporal energy perturbation of the  heating term in Equ.~\ref{eq:energy} could lead to pseudo-periodic behavior of the system with a characteristic time $\tau\sim \tau_\mathrm{rc}$ of a few hours, avoiding a thermal nonequilibrium process and in agreement with our observations.

Going back to the data, we could try to verify the relationship between $\tau_\mathrm{rc}$, $L$, and $T$ given by Equ.~\ref{eq:tau_rc}. Since we currently only have observations in a single passband, we have to assume the plasma temperature. Furthermore, our code does not provide the length of the loops, only the area of the detected regions. Nevertheless, as a first order estimate, we plotted in the bottom right panel in Fig.~\ref{fig:synthesis_histograms} the square root of the area {\it vs.} the period of the events. No correlation is apparent, but there is a region of exclusion. Events with short periods tend to have smaller areas, which, given the unknowns, may be an early indication that the relationship is indeed verified.

\section{Summary and conclusions}

Using EIT 19.5 nm sequences, we detected 917 events of intensity pulsations with periods of 3 to 16 hours. We carefully investigated the possible source of instrumental artifacts and concluded that these events are of solar origin. Many of these events last for several days. Even though the amplitude of the pulsations can be large (up to 100\% intensity variations in some cases), these phenomena have remained largely unnoticed up to now, perhaps because their detection requires very long uninterrupted observations. Out of the 917 events, about half are located in quiet Sun areas, and half in active regions. Of the second group, at least half are associated with coronal loops and their number distribution closely follows solar activity.

The physical cause of these pulsations remains uncertain. We have explored possible explanations for the active region events. Their main characteristics exclude any interpretation in terms of waves. Numerical simulations of thermal nonequilibrium can produce periods consistent with the observed range, but these models still fail to reproduce all the observed properties of loops. From the energy equation of a constant pressure loop, we have showed that we can expect a response time to small perturbations around equilibrium of the order of a few hours. However, we still need to identify what form of heating term would actually produce a cyclic response. 

It is legitimate to ask ourselves if the observed pulsations are common phenomena, in which case their study could be very relevant to active region physics. We estimated the number of distinct active regions during the year 2000, at the maximum of activity cycle 23. There were about 480 regions recorded by the National Oceanic and Atmospheric Administration in 2000. Since large active regions have life times of the order of 1 to 2 months~\citep{Zuccarello2013}, we can estimate the number of distinct regions to be about 240 distinct regions. We detected 100 events in active regions during the year 2000 (central panel in Fig.~\ref{fig:synthesis_histograms}), meaning that on average half of the active regions underwent an episode of intensity pulsations. Since the detection thresholds that we used are rather severe, it is plausible that a more refined search would yield more detections. In the end, we can conclude that these phenomena are quite common, and could actually be the norm. If that is confirmed, the measurement of the properties of the pulsations may provide a new and pertinent diagnostic of the heating mechanism(s) in active regions.

\begin{acknowledgements}
The authors acknowledge the use of the CME catalogue generated and maintained at the CDAW Data Center by NASA and The Catholic University of America in cooperation with the Naval Research Laboratory. The authors acknowledge the use of the Sunspot Number data provided by the Solar Influences Data Analysis Center. SOHO is a project of international cooperation between ESA and NASA.
\end{acknowledgements}

\bibliographystyle{aa}
\bibliography{bibliography}

\appendix

\section{Sources of spurious frequencies\label{sec:biases}}
Great care must be taken when interpreting the frequency content of intensity time series obtained with EIT. Two categories of artifacts can introduce spurious frequencies in the data: instrumental artifacts and geometrical artifacts.

\begin{figure}
\centering
\includegraphics[scale=0.59]{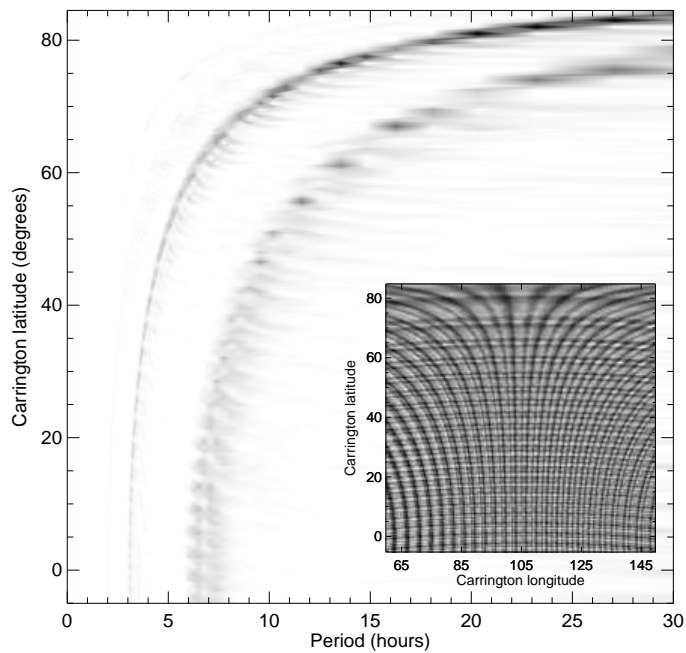}
\caption{Results of the simulation consisting of images containing only the EIT grid pattern. The main image shows Fourier power spectra for points in the middle of the ROI. Darker shades of gray represent larger powers. An example of a simulated image is shown in the lower-right corner.}
\label{fig:simul}
\end{figure}

\subsection{Instrumental artifacts\label{sec:artifacts}}
In EIT, a thin-film (150 nm) aluminium filter is placed on the flange that separates the telescope section from the focal plane assembly, thus preventing visible stray-light from reaching the charge-coupled device (CCD) detector. This filter is supported by a nickel mesh grid (440 $\mu\mathrm{m}$ pitch, 40~$\mu\mathrm{m}$ rods) that casts a square periodical shadow on the CCD~\citep{Defise99, Auchere11}. Because of the strong EUV radiation-induced degradation of the CCD~\citep{Defise99, defise1999b}, a negative image of this constant pattern became part of the flat-field. This effect is well understood and its compensation is part of the standard EIT calibration software included in the IDL Solar Software library. However, slight under- or overcorrections
produce grid residuals in the calibrated images. From visual inspection of the EIT movies, the residuals form a pattern morphologically similar to the original grid, but of reduced amplitude. Even though they never exceed more than a small percentage of the signal, these residuals are clearly visible in the EIT movies because they form a fixed periodical pattern superimposed on the rotating Sun. After mapping into heliographic coordinates, the effect is reversed: the solar surface is stationary, while the flat-field residuals drift from higher to lower longitudes. In the case where the $B_0$ tilt angle of the solar rotation axis is equal to zero, the tiles of the grid pattern pass one after another in front of a given point of the heliographic grid separated in time by
\begin{equation}
\Delta T=\frac{DP\Delta x}{R_{\sun}\Omega(\phi)\cos\phi\cos\theta(t)},
\label{eq:artifacts}
\end{equation}

\noindent
where $D\approx 213\ R_\sun$ is the Sun to {\it SOHO} distance, $P=2.627^{\prime\prime}.\mathrm{pix}^{-1}$ is the plate scale of EIT~\citep{Auchere00, auchere2004}, $\Delta x=21\ \mathrm{pixels}$ is the pitch of the grid, $\Omega(\phi)$ is the rotation rate measured by~\citet{Hortin03}, $R_\sun=696$~Mm is the solar radius, $\phi$ is the latitude of the considered point of the solar surface, and $\theta(t)$ is its Stonyhurst longitude (relative to the central meridian). The period tends towards infinity when the ROI is at the east or west limb ($\theta=\pm 90\degr$), and reaches a minimum of 5.6 hours when the ROI is at the equator ($\phi=0\degr$) and crosses the central meridian ($\theta=0\degr$). Since the period behaves like $1/\cos\theta(t)$, it varies little during most of a half solar rotation. At the equator, the period is inferior to 20 hours 80\% of the time and inferior to 10 hours 60\% of the time. Most of the variation occurs when the ROI is close to the limb. The behavior is similar at higher latitudes, only with larger periods. From this overall stability one can, for example, predict that a period of about 6.2 hours will clearly show up in the Fourier power spectra of structures located at 25\degr of latitude. Furthermore, the grid pattern is not a pure sine modulation but is of composed several components~\citep{clette2002} that can also be revealed by the Fourier analysis.

In order to evaluate more precisely the spurious frequencies caused by the grid residuals, we performed a Fourier analysis on a simulated series of images containing only the grid pattern. The simulation was done using the same programs as for the real data, each original EIT image being simply replaced by the grid modulation pattern. The grid residuals are not identical to the grid pattern itself, and their amplitude varies across the images. However, since they are visually similar (at least locally) to the grid, this simulation should be representative of their effect on the analysis of the real data. In Fig.~\ref{fig:simul} we show the power spectra along the meridian crossing the center of the ROI. Darker shades of gray represent larger powers. The simulated image at the middle of the sequence is inserted in the lower-right corner. The fundamental period (corresponding to the 21 pixels pitch) and the first harmonic (10.5 pixels) are clearly visible. The periods are consistent with the above first-order computation. The fundamental period is about 7 hours at the equator, increases slowly up to 9 hours at mid-latitudes, and then increases more rapidly at higher latitudes. The secondary component of the grid pattern, which has a period of about 10~pixels, forms a thinner peak at half these values. Lower, fainter peaks are also visible at shorter periods. For other heliographic meridians in the ROI, the frequencies are very similar, but the width of the peaks and their relative levels vary. Since the grid residuals are present in every EIT image, these frequencies will show up in any frequency analysis of intensity time series obtained by tracking structures across the disk. So, between 0\degr and 50\degr of latitude, periods between 3 and 10 hours will be present, which overlaps with the range of periods that we detect. However, we can identify spurious detections by identifying those events that follow the latitude {\it vs.} period relationship described above (see top left panel in Fig.~\ref{fig:synthesis_histograms}). In addition, the grid residuals do not represent more than a small percentage of the signal, which is smaller than the amplitude of most detected events. 

\subsection{Geometrical artifacts}
   
We consider in this paper the emission of an optically thin line emitted by a volume of plasma of unknown geometry. The heliographic mapping transformation assumes that the observed emission originates from a thin shell. This is certainly not verified in many cases, because the emission of a number of structures of different types is superimposed along the line of sight. The distortion of the coronal loops is, for example, especially visible when the structures are close to the limb. In the surroundings of an active region, it is not uncommon to see fanned out loops of which only the base is visible. Such a structure can be spatially quasiperiodic. In heliographic coordinates, as they go from the east to the west limb, these rays will appear to be rotating around their base. Therefore, the light curve of a point on the heliographic grid that is swept by this fan will be artificially periodic. We could not clearly identify an example of this effect in the sequences that we analyzed.

\end{document}